\documentclass{PoS} 
\usepackage{epsfig}
\usepackage{color}
\usepackage{amsmath}
\def\CC{{\cal C}}

\def\CO{{\cal O}}

\def\CM{{\cal M}}

\def\MeV{\mathop{\rm MeV}\nolimits}
\def\GeV{\mathop{\rm GeV}\nolimits}

\def\bar{\overline}

\def\tilde{\widetilde}

\def\bk{$B_K\;$}

\def\etal{{\it et al.}}

\def\spose#1{\hbox to 0pt{#1\hss}}
\def\ltapprox{\mathrel{\spose{\lower 3pt\hbox{$\mathchar"218$}}
 \raise 2.0pt\hbox{$\mathchar"13C$}}}
\def\gtapprox{\mathrel{\spose{\lower 3pt\hbox{$\mathchar"218$}}
 \raise 2.0pt\hbox{$\mathchar"13E$}}}
\def\inapprox{\mathrel{\spose{\lower 3pt\hbox{$\mathchar"218$}}
 \raise 2.0pt\hbox{$\mathchar"232$}}}

\PoS{PoS(LAT2005)348}

\title{The Kaon B-Parameter in Staggered Chiral Perturbation Theory}

\ShortTitle{$B_K$ in S$\chi$PT}

\author{\speaker{Ruth S. Van de Water} and Stephen R. Sharpe \\
Physics Department\\
University of Washington\\
Seattle, WA 98195-1560\\
E-mail: \email{ruthv@u.washington.edu},
\email{sharpe@phys.washington.edu}}

\abstract{We calculate the kaon B-parameter, $B_K$, to next-to-leading order in
staggered chiral perturbation theory. We extend the usual power counting
to include the effects of using perturbative (rather than non-perturbative)
matching factors. Taste breaking enters through
the $O(a^2)$ terms in the effective action, through mixing with
higher-dimension operators, and through the truncation of matching factors.
These effects cause mixing with several additional operators,
complicating the chiral and continuum extrapolations.  We summarize the results here;  all details can be found in Ref.~\cite{VandeWater:2005uq}.
}

\FullConference{XXIIIrd International Symposium on Lattice Field Theory\\
		 25-30 July 2005\\
		 Trinity College, Dublin, Ireland}
\begin{document}
%

%
\section{Introduction}
%
\vspace{-2mm}

Experimental measurements of CP violation can be used to extract information about the CKM matrix.  In particular, the size of indirect CP violation in the neutral kaon system, $\epsilon_K$ \cite{Eidelman:2004wy}, combined with theoretical input, places an important constraint on the apex of the CKM unitarity triangle \cite{CKMfit}.  The dominant source of error in this procedure is the uncertainty in the lattice determination of the nonperturbative constant $B_K$, which parameterizes the $K^0-\bar{K^0}$ matrix element.      

Promising calculations with partially quenched staggered fermions are in progress \cite{Gamiz:2004qx,W_Lee}.  Because staggered fermions are computationally cheaper than other standard discretizations, they allow simulations with the lightest dynamical quark masses currently available.  Unfortunately, staggered fermions come with their own source of error -- taste symmetry breaking.  Each lattice fermion flavor comes in four tastes, which are degenerate in the continuum.  The nonzero lattice spacing, $a$, breaks the continuum taste symmetry at $\CO(a^2)$, and the resulting discretization errors are numerically significant at present lattice spacings \cite{MILCf}.  Thus one must use the chiral perturbation theory functional forms for \emph{staggered} fermions in order to correctly perform the combined continuum and chiral extrapolations incorporating taste violations \cite{BA1, Us}.       

\vspace{-3mm}
%
\section{$B_K$ with Staggered Quarks}
%
\vspace{-1mm}

The kaon $B$-parameter is defined as a ratio of matrix elements:
\begin{equation}
	\CM_K = \langle\bar{K^0} | \CO_K | K^0\rangle = B_K \CM_\textrm{vac} , 
\label{eq:BKdef}\end{equation}
where $\CO_K$ is a weak operator and $\CM_\textrm{vac}$ is the result for $\CM_K$ in the vacuum saturation approximation:
\begin{eqnarray}
	\CO_K &=& [\bar{s_a} \gamma_\mu (1 + \gamma_5) d_a][\bar{s_b}\gamma_\mu(1 + \gamma_5) d_b] \\
	\CM_\textrm{vac} &=& \frac{8}{3} \langle \bar{K^0} |[\bar{s_a} \gamma_\mu (1 + \gamma_5) d_a]|0\rangle \langle 0|[\bar{s_b}\gamma_\mu(1 + \gamma_5) d_b]| K^0\rangle . \label{eq:MVac} 
\end{eqnarray}
Note that there are separate summations over color indices $a$ and $b$.  Thus the matrix element $\CM_K$ receives contributions from two different quark-level contractions, one of which produces a single color loop and the other which has two color loops.  

In order to calculate \bk with \emph{staggered} fermions we must introduce the \emph{taste} degree of freedom, both in the operators ($\CO_K$) and in the states ($K^0$ and $\bar{K^0}$).  We choose the external staggered kaons to be taste $P$, by which we mean that the lattice meson operator contains the pseudoscalar taste matrix $\xi_5$.  Since this is the lattice Goldstone taste, its correlation functions satisfy $U(1)_A$ Ward identities, so S$\chi$PT expressions for its mass, decay constant, and other physical quantities are simpler than those for other tastes of PGBs.  In addition, it is a local kaon on the lattice, and therefore relatively simple to implement numerically.  Because the external kaons are taste $P$, the weak operator should also be taste $P$:    
\begin{equation}
	\CO_K^{naive} = \big[\bar{s_a} \big(\gamma_\mu (1 + \gamma_5) \otimes \xi_5\big) d_a\big]\big[\bar{s_b} \big(\gamma_\mu(1 + \gamma_5)\otimes \xi_5 \big) d_b\big]
\end{equation}  
However, when this operator is Fierz-transformed, it mixes with other tastes.  To remedy this, we introduce two types of valence quarks, 1 and 2, into the \bk operator:
\begin{eqnarray}
	\CO_K^{staggered} & = & 2 \Big\{ \big[\bar{s1_a} \big(\gamma_\mu (1 + \gamma_5) \otimes \xi_5\big) d1_a\big]\big[\bar{s2_b} \big(\gamma_\mu(1 + \gamma_5)\otimes \xi_5 \big) d2_b\big] \nonumber \\
	& + & \big[\bar{s1_a} \big(\gamma_\mu (1 + \gamma_5) \otimes \xi_5\big) d1_b\big]\big[\bar{s2_b} \big(\gamma_\mu(1 + \gamma_5)\otimes \xi_5 \big) d2_a\big] \Big\}
\label{eq:OKStag}\end{eqnarray} 
and take the matrix element between two types of kaons:
\begin{equation}
	\CM_K^{staggered} = \langle\bar{K^0_1}_P | \CO_K^{staggered} | {K^0_2}_P\rangle\,,
\label{eq:MKStag}\end{equation}
where $K^0_1$ is an $\bar{s1}d1$ meson and $K^0_2$ is an $\bar{s2}d2$ meson.  The extra valence quarks require explicit inclusion of both color contractions in $\CO_K^{staggered}$, while the overall factor of 2 ensures that $\CM_K$ and $\CM_K^{staggered}$ have the same total number of contractions.  When $\CO_K^{staggered}$ is Fierz-transformed, it has a new \emph{flavor} structure so it does not contribute to $\CM_K^{staggered}$.  Thus $\CM_K^{staggered}$ cannot receive contributions from incorrect tastes. 

Current calculations of \bk with staggered quarks use the partially quenched approximation.  The quark content the corresponding PQ theory is quite large.  We have already seen that staggering requires two sets of $d$ and $s$ valence quarks.  Partial quenching adds two corresponding sets of ghost quarks as well as three sea quarks, resulting in \emph{eleven} total quark flavors, each of which comes in four tastes.  In order to make this completely clear, we show the explicit form of the quark mass matrix:
\begin{equation}
	M = diag\{\underbrace{m_uI,\, m_dI,\, m_sI}_\textrm{sea},\, \underbrace{m_{x}I,\, m_{y}I}_\textrm{valence 1},\, \underbrace{m_{x}I,\, m_{y}I}_\textrm{valence 2},\, \underbrace{m_{x}I,\, m_{y}I}_\textrm{ghost 1},\, \underbrace{m_{x}I,\, m_{y}I}_\textrm{ghost 2}\} ,
\end{equation}
where $I$ is the $4 \times 4$ identity matrix.    

\vspace{-3mm}
%
\section{Generalized S$\chi$PT for $B_K$}
%
\vspace{-1mm}

The expression for \bk in S$\chi$PT must describe \bk at $a\neq0$ if it is to be used for continuum and chiral extrapolations of lattice data; we therefore discuss how \bk is actually calculated on the lattice.  Additional details can be found in Ref.~\cite{VandeWater:2005uq}.  The \bk matrix element, $\CM_K^{staggered}$, receives a contribution from the lattice version of $\CO_K^{staggered}$, as well as from other lattice operators that are in the same representation of the symmetry group that maps a hypercube onto itself \cite{Verstegen}:  
\begin{eqnarray}
	\CO_K^{staggered, cont} &=& \CO_K^{staggered, lat} + \frac{\alpha}{4\pi}[\mbox{taste }P\mbox{ ops.}] + \frac{\alpha}{4\pi}[\mbox{other taste ops.}]\nonumber\\
	&& + \alpha^2[\mbox{all taste ops.}] + a^2[\mbox{all taste ops.}]  + \ldots ,
\label{eq:Mixing}\end{eqnarray}
where $\alpha$ is the strong coupling constant. The 1-loop perturbative matching coefficients between $\CO_K^{staggered}$ in the continuum and four-fermion lattice operators are known, and are numbers of order unity times $\alpha/4\pi$ \cite{LSMatch}.  However, the 2-loop matching coefficients have not been determined, so, in order to remain conservative, we consider them to be of order unity times $\alpha^2$ \emph{without} any factors of $4\pi$.  Current numerical staggered calculations are in fact of the following matrix element:
\begin{equation}
  \langle\bar{K^0_1}_P| \CO_{1-loop}(\mbox{taste }P) |{K^0_2}_P\rangle \equiv \CM_\textrm{lat},
\label{eq:LatOp}\end{equation}
where the subscript ``$1-loop$'' and the argument ``taste $P$'' indicate that one includes all staggered lattice operators with taste $P$ that mix with the latticized \bk operator at $\CO(\alpha/4\pi)$, i.e. those in the second term on the RHS of Eq.~(\ref{eq:Mixing}), using the appropriate matching coefficients.  However, one neglects wrong-taste and higher-order perturbative mixing (terms three and four), as well as all operators which arise through discretization effects (term five), in Eq.~(\ref{eq:Mixing}).  Although the expression in Eq.~(\ref{eq:LatOp}) differs from the continuum matrix element, it reduces to the desired quantity in the continuum limit.  Generically,
\begin{equation}
  \CM_\textrm{lat} = \CM_\textrm{cont} + \frac{\alpha}{4\pi}\CM' + \alpha^2 \CM'' + a^2 \CM''' + \ldots ,
\end{equation}
where $\CM_\textrm{cont}$ is the desired continuum result.  The matrix element $\CM'$ comes from neglecting taste-violating 1-loop operator mixing, while $\CM''$ comes from neglecting 2-loop operator mixing.  Both taste-breaking and taste-conserving discretization errors generate $\CM'''$.  Our goal is therefore to calculate the \emph{lattice} matrix element, $\CM_\textrm{lat}$, to NLO in S$\chi$PT.  One can fit the resulting expression to lattice data, determine the various errors, and remove them, thereby extracting the continuum matrix element,$\CM_\textrm{cont}$.

The appropriate power-counting scheme for calculating $\CM_\textrm{lat}$ in S$\chi$PT must incorporate $a^2$, $\alpha/4\pi$, and $\alpha^2$.  Continuum $\chi$PT is a low-energy expansion in both the pseudo-Goldstone boson (PGB) momentum and the quark masses;  it assumes that $p_{PGB}^2/\Lambda_{\chi}^2 \sim m_{PGB}^2/\Lambda_{\chi}^2$, where $m_{PGB}^2 \propto m_q$ and $\Lambda_\chi \approx 4\pi f_{\pi}$ is the $\chi$PT scale.  However, the numerical values of $m_q$, $a^2$ and $\alpha$ all depend on the particular parameters of a given lattice simulation.  Current PQ staggered lattice simulations \cite{MILCf} use a range of PGB masses from $m_{PGB}^2/\Lambda_\chi^2 = 0.04 - 0.2$, so our S$\chi$PT expression must apply throughout this range.  Generic discretization errors are of the size $a^2 \Lambda_\textrm{QCD}^2$, which is approximately $0.04$ for $1/a \sim 2\GeV$ and $\Lambda_\textrm{QCD} \sim 400 \MeV$, so they are comparable to the minimal $m_{PGB}^2/\Lambda_{\chi}^2$ and should be included at the same order.  Taste-breaking discretization errors, on the other hand, are caused by exchange of gluons with momentum $\pi/a$, and therefore receive an additional factor of $\alpha^2_V(\pi/a)$, so their size must be considered separately.  At the lightest quark masses, the lattice Goldstone pion mass is comparable to the mass splittings among the other PGB tastes: $m_{PGB}^2/\Lambda_{\chi}^2 \sim a^2 \alpha_V^2(q^* = \pi/a) \Lambda^2 \sim 0.04$, where $\Lambda$ is a QCD scale which turns out to be around $1200 \MeV$.  Thus they are not suppressed relative to pure $\CO(a^2)$ discretization effects by the additional powers of $\alpha$, as naive power-counting would suggest, because of the large scale $\Lambda$ associated with the taste-breaking process at the quark level \cite{Mason:2002mm}.  Standard S$\chi$PT only includes discretization effects -- we now consider additional errors from perturbative operator matching, which depend upon $\alpha_V(q^*)$.  Generically, the choice of $q^*$ is process-dependent, and the value of $\alpha_V(q^*)$ ranges from $\sim 0.3 - 0.55$ for $q^* = \pi/a - 1/a$ at $a=.125$fm \cite{CB_alpha}.  Thus both $\alpha/4\pi$ and $\alpha^2$ must be included at lowest-order in our power counting.

In light of this discussion, we adopt the following extended S$\chi$PT power-counting scheme:      
\begin{equation}
	p^2 \sim m \sim a^2 \sim a_\alpha^2 \sim \alpha/4\pi \sim \alpha^2\,,
\end{equation}
where $a_\alpha^2 \equiv \alpha^2_V(\pi/a) a^2$.  We account for the fact that $\alpha^2 a^2$ terms in the action are enhanced numerically by including them at the same order as simple discretization effects.  In fact, while it may seem ad hoc, $p^2 \sim m \sim a_\alpha^2$ is the \emph{standard} S$\chi$PT power counting scheme.  It is simply not traditionally written as such because standard S$\chi$PT calculations have only included $\CO(a_\alpha^2)$ taste-breaking discretization errors from the action, and have therefore not needed to contrast them with pure $\CO(a^2)$ discretization effects.  We also use conservative power-counting for the perturbative errors by assuming that 2-loop contributions are not significantly smaller than those from 1-loop diagrams.  We emphasize that our scheme is \emph{phenomenologically based} on the particular parameter values of current staggered simulations -- simulations using significantly lighter quark masses or smaller lattice spacings would require a different scheme and result in a different S$\chi$PT expression for \bk.   

In S$\chi$PT, $\CM_\textrm{lat}$ is simply the matrix element of a sum of operators,
\begin{equation}
  \CM_\textrm{lat} = \langle \bar{K^0_1}_P | \CC_\chi^i \CO_\chi^i | {K^0_2}_P \rangle, 
\end{equation}
with undetermined coefficients.  In order to calculate $\CM_\textrm{lat}$ at NLO we must include operators in the chiral effective theory of the following order:
\begin{eqnarray}
  \CO(p^2), \;\CO(m), \;\CO(a^2), \;\CO(a_\alpha^2), \;\CO(\alpha/4\pi), \;\CO(\alpha^2) \,.
\end{eqnarray}
We determine all of the linearly-independent operators of this order using the graded group theory method of Ref.~\cite{Me} along with the taste spurion method of Refs.~\cite{LS,Us}.  Directly mapping $O_K^{staggered}$ onto chiral operators leads to operators of of $\CO(p^2)$, truncation of perturbative matching factors leads to operators of $\CO(\alpha/4\pi)$,$\CO(\alpha^2)$, generic discretization effects lead to operators of $\CO(a^2)$, and insertions of taste-breaking operators from the staggered action with $O_K^{staggered}$ lead to operators of $\CO(a_\alpha^2)$.  In total, we find thirteen chiral operators;  they are given in Eq.~(27) of Ref.~\cite{VandeWater:2005uq}.  We note, however, that many of these chiral operators arise in more than one way and correspond to more than one quark level operator, so they actually have more than one undetermined coefficient.

\vspace{-4mm}
%
\section{Results and Conclusions}
%
\vspace{-2mm}

We have calculated \bk to NLO in S$\chi$PT  for quenched, PQ, and full QCD;  the results are in Ref.~\cite{VandeWater:2005uq}.  The simplest expression that still shows the general structure of \bk in S$\chi$PT is the partially quenched expression for for three sea quarks and degenerate valence quarks ($m_x = m_y$):  
\begin{eqnarray}
  B_K &=& B_0\Bigg\{1 + \frac{1}{512 \pi^2 f_{K_P}^2} \sum_{B'} f^{B'} \bigg( \ell(m^2_{K_{B'}}) - \frac{1}{2} m^2_{K_{B'}} \tilde\ell(m^2_{K_{B'}}) \bigg) + A \Bigg\} \nonumber\\
    &+& B\frac{3m^2_{K_P}}{16f_{K_P}^2} + D\frac{3(m_u + m_d + m_s )}{16f_{K_P}^2}\nonumber\\
  &-&  \,\frac{3}{512 \pi^2 f_{K_P}^2} \sum_{B'} \tilde\ell(m^2_{K_{B'}}) \sum_B \left(\frac{\CC^{1B}_\chi}{f^4} g^{BB'} - \frac{\CC^{2B}_\chi}{f^4} h^{BB'} \right) \,. 
\label{eq:BKNLO}\end{eqnarray}
The functions $\ell$ and $\tilde\ell$ are chiral logarithms, and the index $B'$ runs over the PGB tastes $I, P, V, A$, and $T$.  The fact that all PGB tastes enter through loops is a generic trait of S$\chi$PT.  The 1-loop contribution in Eq.~(\ref{eq:BKNLO}) proportional to $B_0$ and the analytic terms proportional to the coefficients $B$ and $C$ reproduce the continuum $\chi$PT result when $a \rightarrow 0$ \cite{BSW}.  The 1-loop contributions proportional to the coefficients $\CC^{1B}$ and $\CC^{2B}$ are new, and come perturbative matching and discretization errors.  Because they have a different functional form from the continuum 1-loop expression, they can, in principle, be determined and removed separately at each lattice spacing.  There remains, however, a multiplicative correction to $B_0$ from the discretization and perturbative matching errors, $A$, which can only be removed using a fit to multiple lattice spacings.   

Because of the large number of fit parameters and complexity of the expression, we outline a fitting strategy that takes advantage of other matrix elements and maximizes information that can be extracted from a single lattice spacing in Ref.~\cite{VandeWater:2005uq}.  Even with the implementation of these suggestions, however, fitting the staggered \bk data will be highly nontrivial.  One would clearly like to somehow decrease the number of operators, or at least undetermined coefficients, that contribute to \bk at NLO.  One way to do this is by using improved links.  This drastically reduces the 1-loop perturbative mixing between $\CO_{1-loop}(\textrm{taste P})$ and certain wrong-taste operators.  The biggest source of potential improvement, however, is in better perturbative matching.  A good first step would therefore be to match to \emph{all tastes of lattice operators} at 1-loop, since the requisite matching coefficients are known.  While it has been thought that ``wrong taste'' operator contributions to \bk would be less important than those from taste $P$ operators, our power-counting and operator enumeration show that this is not the case in general.  Fully nonperturbative matching, although extremely difficult, would completely eliminate some of the thirteen operators.     

$B_K$ is an important parameter for constraining the phase of the CKM matrix, and consequently new physics.  Calculations of $B_K$ with staggered fermions are promising, but they need the appropriate S$\chi$PT form for correct continuum and chiral extrapolations.  We have calculated $B_K$ to NLO in S$\chi$PT for quenched, PQ, and full QCD using an extended power-counting scheme.  The result is \emph{more complicated} than in previous S$\chi$PT cases.  While in expressions for quantities such as $f_K$ and $f_D$, the primary effect of staggering is additive corrections to PGB masses inside loops, the expression for \bk contains entirely new contributions with different functional forms than the continuum piece.  Use of our expression for \bk at NLO in S$\chi$PT, in combination with sufficient lattice data, should allow a precise determination of \bk with staggered quarks.

\vspace{-3mm}
%
%



\end{document}